# The combinatorial multitude of fatty acids can be described by Fibonacci numbers

Stefan Schuster

Dept. of Bioinformatics, Friedrich Schiller University
Ernst-Abbe-Platz 2, 07743 Jena, Germany
Phone +49-3641-949580, Fax +49-3641-946452
Email: stefan.schu@uni-jena.de

**Abstract**

The famous series of Fibonacci numbers is defined by a recursive equation saying that each number is the sum of its two predecessors, with the initial condition that the first two numbers are equal to unity. Here, we show that the numbers of fatty acids (straight-chain aliphatic monocarboxylic acids) with *n* carbon atoms is exactly given by the Fibonacci numbers. Thus, by investing one more carbon atom into extending a fatty acid, an organism can increase the variability of the fatty acids approximately by the factor of the Golden section, 1.618. As the Fibonacci series grows asymptotically exponentially, our results are in line with combinatorial complexity found generally in biology. We also outline potential extensions of the calculations to modified (e.g., hydroxylated) fatty acids. The presented enumeration method may be of interest for lipidomics, combinatorial chemistry, synthetic biology and the theory of evolution (including prebiotic evolution).

**1. Introduction**

Fatty acids are basic components of lipids in all self-replicating organisms. They occur within triglycerides, which serve as energy and carbon stores, and within phospholipids, which serve as main constituents of biomembranes (cf. Nelson and Cox, 2000; Berg et al., 2002). It is well known that a enormous multitude of different fatty acids occur in living organisms, such as stearic acid, palmitoic acid, linoleic acid and many others. They differ in the number of carbon atoms and the number and position of double bonds. When only single bonds occur in the side chain, the fatty acids are called saturated. If at least one double bond occurs, they are called unsaturated. Two unsaturated fatty acids are unconditionally essential constituents of the human diet: alpha-linolenic acid and linoleic acid (cf. Nelson and Cox, 2000; Berg et al., 2002). In the case of several double bonds, the term poly-unsaturated fatty acids (PUFAs) is used. An example of a fatty acid with five double bonds is cis-5,8,11,14,17-eicosapentaenoic acid (Bostock et al., 1980).

Fatty acids also act as signalling molecules in the interaction between plants and plant pathogens, and elicit the synthesis of stress metabolites in the plant (Bostock et al., 1980). Obviously, a high specificity is needed in this interaction, promoting a high diversity of fatty acids.

It is not immediately clear whether all theoretically possible fatty acids are really used in living nature. To tackle this question and to elucidate the combinatorial complexity of fatty acids, we here derive a formula for enumerating them in dependence on the number of carbon atoms involved. We derive both a recursion formula and an explicit formula. Finally, we discuss biological implications of the calculations.

Most fatty acids in living organisms include an even number of carbon atoms and are, thus, called even-chain fatty acids (cf. Berg et al., 2002). Odd-chain fatty acids occur in some plants, marine organisms and also in cow milk (cf. Voet and Voet, 2004). An example is provided by pentadecanoic acid, a saturated FA with $n = 15$, which is found at the level of 1.2% in the milk fat from cows (Smedman et al., 1999). Even-chain fatty acids are synthesised by condensing several acetyl-CoA molecules, which represent two-carbon units. Even-chain and odd-chain fatty acids behave differently with respect to the convertibility into sugar (de Figueiredo et al., 2009).

## 2. Fatty acids and Fibonacci numbers

*2.1. Basic assumptions*

First, it should be said that the term "fatty acid" (henceforth abbreviated by FA) is not clearly defined. A common denominator is that each FA involves one carboxy group (hence their acidic character) and a hydrocarbon side chain. In other words, the side chain is composed of methyl groups ($CH_3$), methylene groups ($CH_2$), and/or CH groups with one hydrogen only. The carbons are linked by single or double bonds, with the special feature that never two double bonds are adjacent to each other, which is probably due to the dehydrogenation mechanism in the conversion of saturated into unsaturated FAs.

In counting the number of carbons in FAs, the carbon in the carboxy group is included. Thus, an $n$-carbon FA has $n-1$ carbons in the side chain. Often, only molecules with at least 14 or 16 carbon atoms are considered to be FAs, because within triglycerides and phospholipids, those longer acids are predominantly used. Other authors use the definition less strictly by including all possible chain lengths.

Some of the carbons can be hydroxylated such as in hydroxybutyric acid and in ricinoleic acid, which is an unsaturated FA with 18 carbons and one hydroxy group (James et al., 1965). While the vast majority of FAs involve unbranched side chains, there are some examples of branched FAs such as phytanic acid, which results from degradation of chlorophyll (Brink and Wanders, 2006). Alternatively, this can be considered as an FA modified by methylation at several positions.

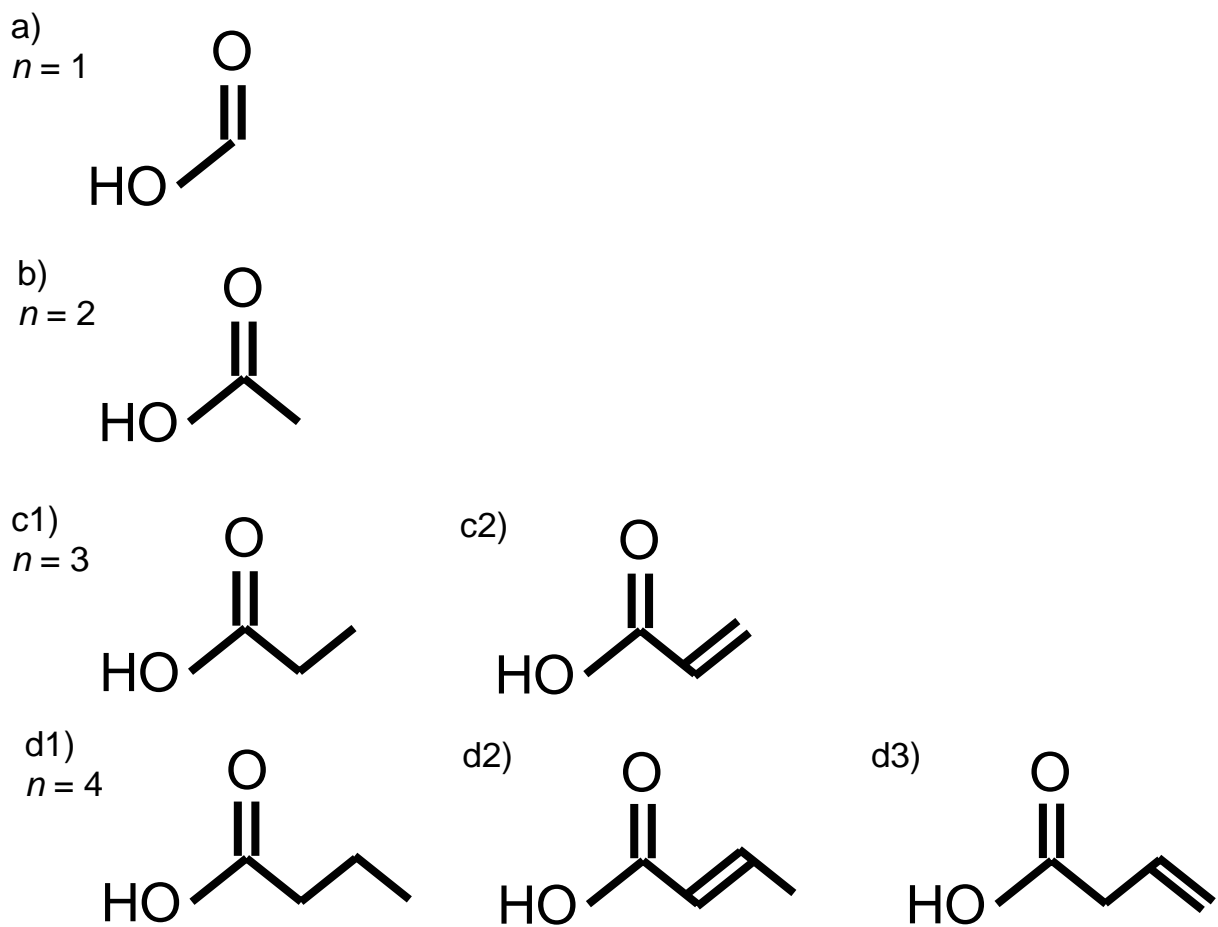

**Fig. 1.** Aliphatic monocarboxylic acids for $n = 1$ up to $n = 4$ in a chemical notation where only the carbon skeleton and functional groups are shown, while hydrogens bonded to carbons are omitted. Trivial names of acids: a) formic acid; b) acetic acid; c1) propionic acid; c2) acrylic acid; d1) butyric acid; d2) crotonic acid; d3) 3-butenoic acid.

Analysing the complete combinatorics of FAs in their widest definition is next to impossible. However, if we restrict the analysis to certain classes of FAs, combinatorial analysis is feasible. First, we use the following restriction: We consider all molecules only consisting of one carboxy group at the upper end and a straight chain of carbons linked by single or double bonds, with the special feature that never two double bonds are adjacent to each other (Fig. 1).

This also implies that the first and second carbons must be linked by single bond, because the first carbon is involved in a double already (in the carboxy group). This is particularly clear because two double bonds would prevent the necessary bond to the OH in the carboxy group.

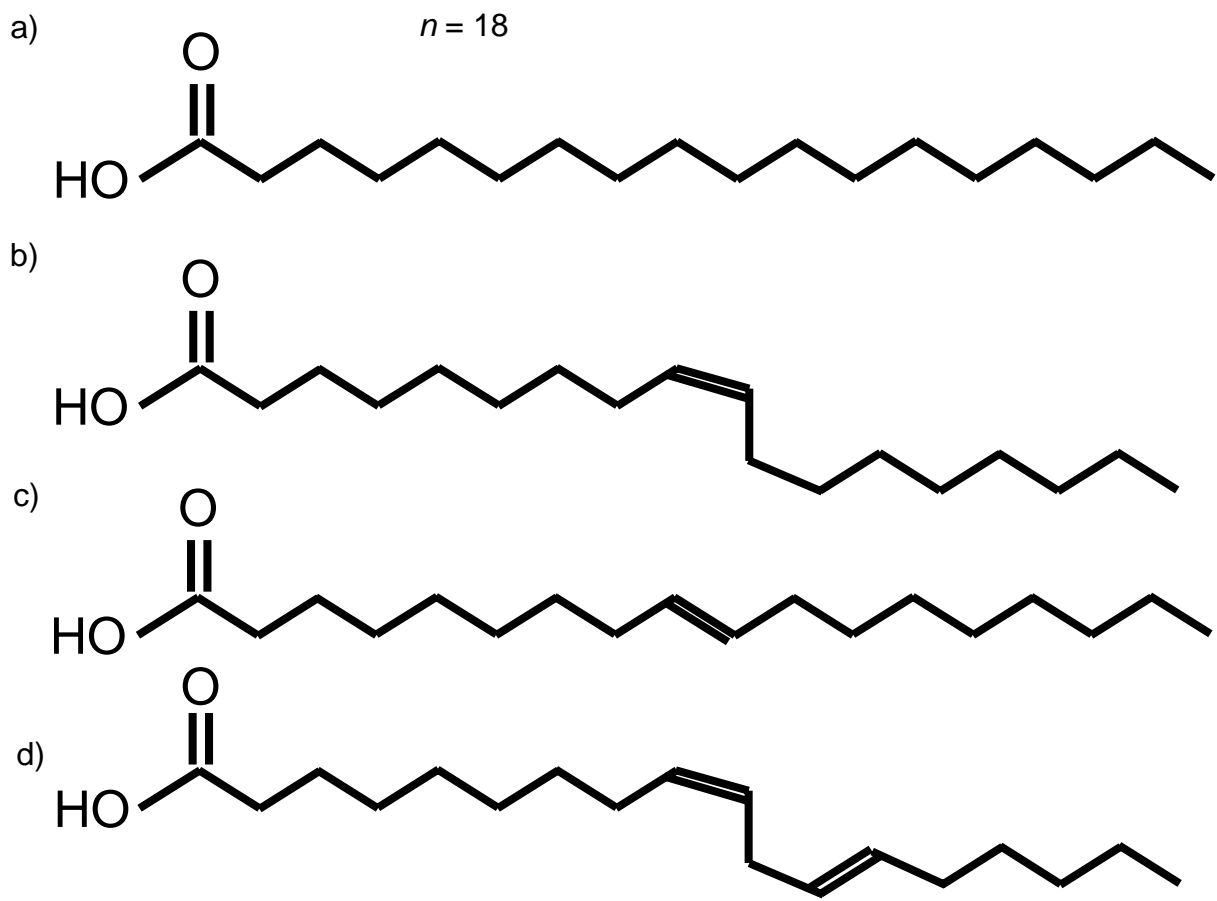

**Fig. 2.** The three major physiological fatty acids with $n = 18$. a) Palmitoic acid (saturated), b) oleic acid (cis-isomer), c) elaidic acid (trans-isomer), d) linoleic acid. Cis- and trans-isomers are not counted separately in the enumeration method presented here.

All remaining valences of all carbons in the chain are bonded to hydrogens. Chemically speaking, this definition includes all straight-chain aliphatic monocarboxylic acids of arbitrary chain length. As the two single bonds adjacent to a double bond can be situated at the same side and at opposite sides, unsaturated FAs occur as cis- or trans-isomers (Fig. 2). These isomers are not, however, counted separately here.
In Section 3, we somewhat widen this definition by allowing for hydroxy and ketone groups in the side chain. When hydroxy groups are included, also stereoisomers can occur. These are not counted separately here either.

## 2.2. Deriving a recursion formula

Let $x_n$ denote the number of theoretically possible acids according to the above definition. Starting with $n = 1$, we just have the carboxy group linked to one hydrogen, which makes up formic acid (HCOOH). For $n = 2$, there is again only one possibility, notably to attach a methyl group to the carboxy group, giving rise to acetic acid (CH$_3$COOH). Thus, $x_1 = 1$ and $x_2 = 1$. For $n = 3$, the saturated FA is called propionic acid. However, there is also the possibility to insert a single bond, between the second and third carbon, giving rise to acrylic acid (Fig. 1). Thus, $x_3 = 2$.

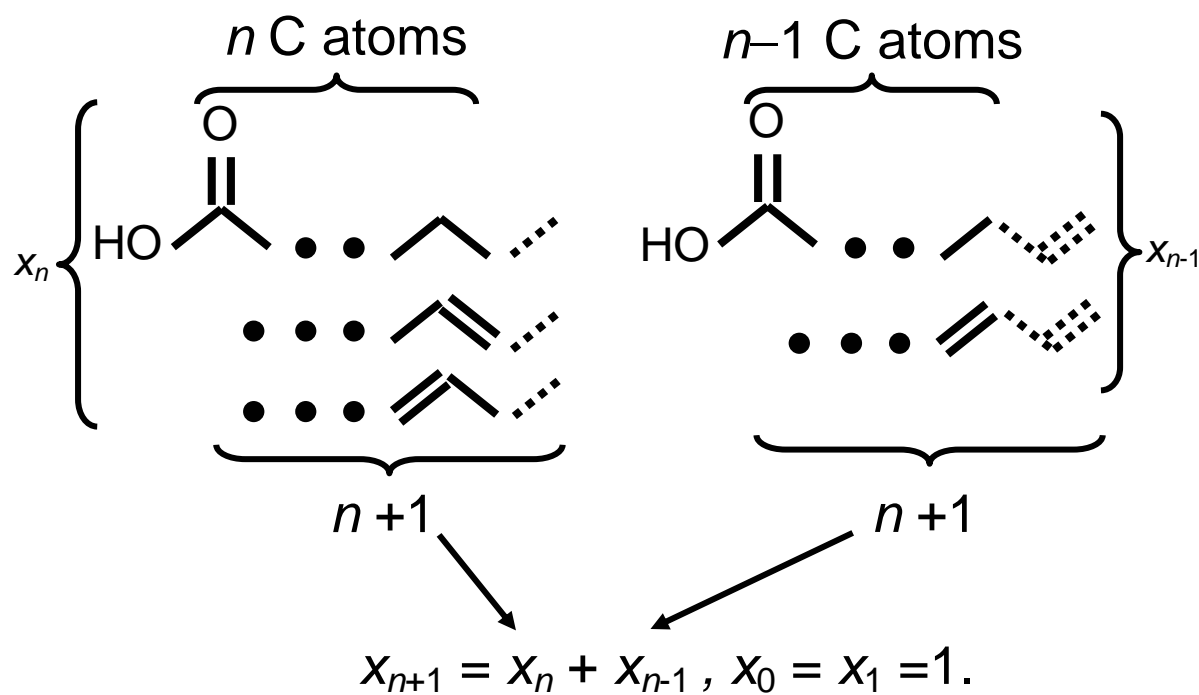

**Fig. 3.** Illustration of the recursive enumeration method. Bonds added during the procedure are dotted. Larger solid dots stand for the remaining chain.

A general enumeration procedure can be derived as follows (Fig. 3). Assume we know all $x_k$ from $k = 1$ up to $k = n$ and wish to calculate $x_{n+1}$. We can certainly extend the molecule by adding one carbon linked to the $n$-th carbon by a single bond. It is not clear, though, whether we can add one carbon linked to the $n$-th carbon by a double bond because it might be that the $n$-th carbon is linked to the $(n-1)$th carbon by a double bond already. To circumvent this problem, we can go back to the molecule with $k = n-1$. To that, we can add two carbon atoms such that carbons $n-1$ and $n$ are linked by a single bond and carbons $n$ and $n+1$, by a double

bond. Combining the two procedures (starting at *n*−1 and at *n*), we arrive at the recursion formula

$$x_{n+1} = x_n + x_{n-1} \tag{1}$$

(see Fig. 2). There is no overlap between the molecules thus generated because the $x_n$ molecules generated by starting from length *n* have a single bond at the lower end, while the $x_{n-1}$ molecules generated by starting from length *n*−1 have a double bond at the lower end. Moreover, all possibilities of extending the molecules according to the defined rules are covered.

**Table 1.** Fibonacci numbers, $x_n$, Pell (2-Fibonacci) numbers, $y_n$, and 3-Fibonacci numbers, $z_n$, for *n* =1–10 and 16–22.

| *n* | $x_n$ | $y_n$ | $z_n$ |
|---|---|---|---|
| 1 | 1 | 1 | 1 |
| 2 | 1 | 2 | 3 |
| 3 | 2 | 5 | 10 |
| 4 | 3 | 12 | 33 |
| 5 | 5 | 29 | 109 |
| 6 | 8 | 70 | 360 |
| 7 | 13 | 169 | 1.189 |
| 8 | 21 | 408 | 3.927 |
| 9 | 34 | 985 | 12.970 |
| 10 | 54 | 2.378 | 42.837 |
| 16 | 987 | 470.832 | 55.602.393 |
| 17 | 1.597 | 1.136.689 | $1.836 \times 10^8$ |
| 18 | 2.584 | 2.744.210 | $6.065 \times 10^8$ |
| 19 | 4.181 | 6.625.109 | $2.003 \times 10^9$ |
| 20 | 6.765 | 15.994.428 | $6.616 \times 10^9$ |
| 21 | 10.946 | 38.613.965 | $2.185 \times 10^{10}$ |
| 22 | 17.711 | 93.222.358 | $7.217 \times 10^{10}$ |

Eq. (1) is the recursion formula generating the famous series of Fibonacci numbers with the initial conditions $x_1 = 1$ and $x_2 = 1$ (cf. Nishino, 1987; Jean, 1994; Falcón and Plaza, 2007). Coincidentally, we have the same initial conditions here. The series reads

$$x_n = 1, 1, 2, 3, 5, 8, 13, 21, \ldots \tag{2}$$

For example, the three FAs for $n = 4$ are butanoic acid (saturated), crotonic acid (double bond between second and third carbons) and 3-butenoic acid (double bond at the lower end). Table 1 shows the Fibonacci numbers for $n = 1-10$ and for $16-22$. The latter numbers correspond to the most widely occurring FAs.

*2.3. An explicit formula*

Eq. (1) is a linear recursion formula. The usual solution procedure for this type of equations is to use an exponential function

$$x_n = a\lambda^n \tag{3}$$

Substituting this into the recursion formula (1) leads to the quadratic equation

$$\lambda^2 - \lambda - 1 = 0 \tag{4}$$

with the solutions

$$\lambda_{1/2} = \frac{1 \pm \sqrt{5}}{2}. \tag{5}$$

Interestingly, these two numbers represent the smaller and larger ratios of the Golden ratio (cf. Jean, 1994; Falcón and Plaza, 2007). The explicit formula is obtained by a linear combination of two exponential functions with the two bases given in Eq. (5).

$$x_n = a_1 \left(\frac{1+\sqrt{5}}{2}\right)^n + a_2 \left(\frac{1-\sqrt{5}}{2}\right)^n \tag{6}$$

The coefficients $a_1$ and $a_2$ are determined by using the initial conditions. It is convenient to start with $n = 0$ rather than $n = 1$ because any number to the power of zero gives unity. $x_0$ is obtained as $x_0 = x_2 - x_1 = 0$. Thus, Eq. (6) gives, for $n = 0$ and $n = 1$:

$$0 = a_1 + a_2, \quad 1 = \frac{a_1 + a_2}{2} + \frac{a_1 - a_2}{2}\sqrt{5} \tag{7a,b}$$

This leads to the explicit formula

$$x_n = \frac{1}{\sqrt{5}}\left(\frac{1+\sqrt{5}}{2}\right)^n - \frac{1}{\sqrt{5}}\left(\frac{1-\sqrt{5}}{2}\right)^n \tag{8}$$

Although this formula involves irrational numbers, the resulting numbers are integers. This is because the factor $\sqrt{5}$ in the numerators and denominators cancels out when calculating the particular $x_n$.

It is known from mathematics that the ratio of two consecutive Fibonacci numbers tends to the Golden section. This can be shown by substituting Eq. (8) into $x_{n+1}/x_n$. As the modulus (absolute value) of the minus solution in Eq. (5) is smaller than that of the plus solution, it can be neglected in that ratio for large $n$. This leads to the following observation. In the construction procedure of the FAs shown in Fig. 3, which we used to show that the recursion formula (1) applies, we added a terminal double bond by starting from the FAs with $n-1$ carbons, while we added a terminal single bond by starting from the FAs with $n$ carbons. As also $x_n/x_{n-1}$ tends to the Golden section, the numbers of FAs with a terminal single bond and a terminal double bond (for a given chain length) are approximately in the ratio of the Golden section, 1.618. The inverse ratio (FAs with terminal double bond to FAs with terminal single bond) is 0.618. Note that the digits after the period are the same, which is one of the striking properties of the Golden section.

### 3. Possible extensions

The above calculations can be extended for various cases of modified FAs. For example, when ketone (oxo) groups are allowed, one or several carbons can be linked with oxygen atoms by double bonds. An example is acetoacetic acid ($n = 4$). Since carbons are of valence four and the hydrocarbon chain must not be interrupted, ketone groups are not allowed to be adjacent to double bonds within the hydrocarbon chain. Therefore, the recursion procedure outlined in Section 2.2. and illustrated in Fig. 2 can be adapted by saying that a ketone group can be appended by starting from the FA with $n$ carbons, adding a single bond to a carbon involved in a ketone group. When adding a double bond between two carbons, we again start at the FA with $n-1$ carbons, but cannot then add a ketone group. This leads to the recursion formula

$$y_{n+1} = 2y_n + y_{n-1} \tag{9}$$

and the initial conditions $y_1 = 1$, $y_2 = 2$. Note that, in contrast to the Fibonacci numbers, $x_2 = 2$ because a ketone group can occur already in a two-carbon acid. This is glyoxylic acid, after which the glyoxylate cycle is called (cf. Nelson and Cox, 2000; de Figueiredo et al., 2009). Together with these initial conditions, Eq. (9) leads to the series

$$y_n = 1, 2, 5, 12, 29, 70,\ldots \tag{10}$$

(see also Table 1). In mathematics, they are known as the Pell numbers (Bicknell, 1975) or 2-Fibonacci numbers (Falcón and Plaza, 2007). Using again an exponential function ansatz (Eq. (3)) leads to the quadratic equation

$$\lambda^2 - 2\lambda - 1 = 0 \tag{11}$$

with the solution

$$\lambda_{1/2} = 1 \pm \sqrt{2} \tag{12}$$

Taking into account the initial conditions, we obtain the explicit formula

$$y_n = \frac{(1+\sqrt{2})^n - (1-\sqrt{2})^n}{2\sqrt{2}} \tag{13}$$

The same series and formula applies to the case where hydroxy groups are allowed instead of ketone groups, under the restriction that no hydroxy group should be adjacent to a double bond. Such a restriction is reasonable in view of the biosynthesis mechanism of hydroxylated FAs.

When both ketone and hydroxy groups are allowed, we obtain the recursion formula

$$z_{n+1} = 3z_n + z_{n-1} \tag{14}$$

and the initial conditions $z_1 = 1$, $z_2 = 3$. This leads to the series

$$z_n = 1, 3, 10, 33, 109, 360,\ldots \tag{15}$$

and the explicit formula

$$z_n = \frac{1}{\sqrt{13}}\left(\frac{3+\sqrt{13}}{2}\right)^n - \frac{1}{\sqrt{13}}\left(\frac{3-\sqrt{13}}{2}\right)^n \tag{16}$$

In mathematics, they are called 3-Fibonacci numbers (Falcón and Plaza, 2007).

Of course, combinatorial complexity increases with an increasing number of possible functional groups involved. When hydroxy groups are included, also stereoisomers (R and S forms) can occur because some carbons then may have four different binding partners: OH, hydrogen, upper and lower ends of the chain. This as well as including cis- and trans-isomers of unsaturated FAs is an interesing possible extension for future studies.

**Discussion**

Fibonacci numbers are named after Leonardo Pisano, called Fibonacci. In 1202, he published the book "*Liber abaci*", in which he derived this series by studying the population dynamics of rabbits. For a sketch of his curriculum vitae, see Milicević and Trinajstić (2006). The Fibonacci series occurs in many situations in biology such as in phyllotaxis (Jean, 1994) and in secondary structures of proteins (Jean, 1994). It also plays a role in the enumeration of nanotubes composed of zigzag-shaped phenanthrenes (Lukovits and Janezić, 2004; cf. Milicević and Trinajstić, 2006). In physics, it is used, for example, for constructing dielectric multilayers (Gellerman et al., 1994). Interestingly, that method is also employed for measuring the properties of FAs (Ganguly et al., 1997). Moreover, it is used in mathematics for various purposes such as finding extremum values in a given interval (Fibonacci search) (cf. Nishino, 1987). To the best of our knowledge, the observation that the Fibonacci numbers also describe the combinatorial multitude of fatty acids has not been published before.

It is an interesting question which of the theoretically possible FAs are really used in living organisms. It is beyond the scope of this paper to cope with this question. Here, we only mention that an impressive number is really used, but certainly not all FAs. For example, the Fibonacci number for $n = 18$ is $x_n = 2584$ (without counting the derivatives including ketone or hydroxy groups, Table 1). At least four FAs with $n = 18$ play a major role in biology: palmitoic acid acid (saturated), oleic acid and its trans-isomer elaidic acid ($n = 18$, one double bond between carbons 9 and 10, these two are here counted as one) and linoleic acid (two double bonds between carbons 9 and 10 and between carbons 12 and 13) (Fig. 2). Oleic acid has the interesting symmetry property that the double bond is exactly in the middle of the carbon skeleton. Thus, it does not matter whether we count until the 9th carbon from one end or the other.

As the ratio of two consecutive Fibonacci numbers tends to the Golden section, we can derive the following result. Starting from an FA of a given chain length and investing one more carbon atom, an organism can increase the variability of the FA approximately by the factor of the Golden section, 1.618. An interesting question is why most FAs used in living organisms have chain lengths of 16–20, since even longer FAs would provide much more combinatorial complexity. One physico-chemical constraint may arise from the melting temperature. Very long FAs have high melting temperatures and, thus, might be too rigid to be used in organisms.

Beside the academic interest, a promising field of application of this mathematical analysis is lipidomics, that is, the high-throughput detection and measurement of lipids and their constituents, for example, by mas spectrometry. In analysing the mass spectra, it is very helpful to know the maximum number of compounds that can potentially appear.

A further application is to estimate the time necessary to perform, in the laboratory, the chemical synthesis of all FAs of a certain length. This could also be of interest in the emerging discipline of synthetic biology, which is aimed at constructing systems (e.g. gene circuits or metabolic pathways) that have not been present within living organisms earlier (cf. Szostak et al., 2001; Lu et al., 2008). Such engineered systems could produce FAs not found before.

Furthermore, the analysis can help in understanding basic principles of evolution. In many domains of biology, only relatively few building blocks out of an enormous number of theoretically possibilities are used. For example, out of more than 100 chemical elements, only six are mainly used in living organisms: carbon, hydrogen, oxygen, nitrogen, sulphur and phosphorus (cf. Lodish et al. 2000). Only four nucleobases appear in the DNA: guanine, adenine, cytosine and thymine. Proteins are built from a limited set of amino acids. For example, the number of possible aliphatic amino acids, for which a recursion formula can be given (Grützmann et al., 2010) by far exceeds the number of proteinogenic aliphatic amino acids. Biological complexity then arises by a versatile combination of the building blocks.

As for FAs, a higher number really occurs than for nucleobases and amino acids. This might be so because FAs are, in a sense, building blocks and polymers at the same time. Due to the synthesis by assembling two-carbon units (in the form of acetyl-CoA), many possibilities arise in inserting double bonds, hydroxyl groups and other groups. An additional source of complexity is the combination of FAs into phospholipids and triglycerides. This makes the mathematical modelling of lipid biosynthesis a difficult task (cf. Yetukuri et al., 2007; Kenanov et al., 2010).

The calculations presented here may also be of interest in the context of prebiotic evolution. Current theories assume that RNA molecules divided within lipid vesicles (Szostak et al., 2001). Some authors have even put forward the idea of a lipid world in prebiotic evolution (Segrè et al., 2001).

**Acknowledgments**

The author would like to thank Severin Sasso (Jena) and the members of the Dept. of Bioinformatics at the University of Jena, in particular Christina Glock, Konrad Grützmann, Martin Pohl, Heiko Stark and Silvio Waschina, for very helpful discussions. Ina Weiss from this Department helped with the literature search. Financial support by the German Ministry for Education and Research within the Virtual Liver Network is gratefully acknowledged.

**References**

J.M. Berg, J.L. Tymoczko, L. Stryer: Biochemistry, 5th edn., Freeman, New York 2002.

M. Bicknell: A primer on the Pell sequence and related sequences. *Fibonacci Quart.* 13 (1975) 345–349.

R.M. Bostock, J.A. Kuc, R.A. Laine: Eicosapentaenoic and arachidonic acids from *Phytophthora infestans* elicit fungitoxic sesquiterpenes in the potato. *Science* 212 (1981) 67–69.

D.M. Brink, R.J.A. Wanders: Phytanic acid: production from phytol, its breakdown and role in human disease. *Cell. Mol. Life Sci.* **63** (2006) 1752–1765.

L.F. de Figueiredo, S. Schuster, C. Kaleta, D.A. Fell. Can sugars be produced from fatty acids? A test case for pathway analysis tools. *Bioinformatics* 25 (2009) 152–158

S. Falcón, A. Plaza, The *k*-Fibonacci sequence and the Pascal 2-triangle. *Chaos, Solit. Fract.* 33 (2007) 38–49

P. Ganguly, M. Sastry, S. Choudhury, D.V. Paranjape: "Turnover" of amphiphile molecules in Langmuir Blodgett films of salts of fatty acids: An X-ray diffraction study. *Langmuir* 13 (1997) 6582–6588.

W. Gellerman, M. Kohmoto, B. Sutherland, P.C. Taylor: Localization of light waves in Fibonacci dielectric multilayers. *Phys. Rev. Lett.* 72 (1994) 633–636.

K. Grützmann, S. Böcker, S. Schuster: Combinatorics of aliphatic amino acids. *Naturwissenschaften* 98 (2011) 79–86.

A.T. James, H.C. Hadaway, J.P. Webb: The biosynthesis of ricinoleic acid. *Biochem. J.* 95 (1965) 448–452.

R.J. Jean: Phyllotaxis. A systemic study in plant morphogenesis. Cambridge University Press, Cambridge 1994.

D. Kenanov, C. Kaleta, A. Petzold, C. Hoischen, S. Diekmann, R. Siddiqui, S. Schuster: Theoretical study of lipid biosynthesis in wild type *Escherichia coli* and in a protoplast-type L-form using elementary flux mode analysis. *FEBS J.* 277 (2010) 1023–1034.

D.L. Nelson, M.M. Cox: Lehninger Principles of Biochemistry. Worth Publishers, New York 2000.

H. Lodish, A. Berk, S.L. Zipursky, P. Matsudaira, D. Baltimore, J. Darnell: Molecular Cell Biology, 4th edn. Freeman, New York 2000.

X. Lu, H. Vora, C. Khosla: Overproduction of free fatty acids in *E. coli*: implications for biodiesel production. *Metab. Eng.* 10 (2008) 333–339.

I. Lukovits, D. Janezić, Enumeration of conjugated circuits in nanotubes. *J. Chem. Inf. Comput. Sci.* 44 (2004) 410–414.

A. Milicević, N. Trinajstić: Combinatorial enumeration in chemistry. *Chem. Modell.* 4 (2006) 405–469.

H. Nishino: Binary search revisited: Another advantage of Fibonacci search, *IEEE Trans. Computers* C-36 (1987) 1132–1135.

D. Segrè, D. Ben-Eli, D. Deamer, D. Lancet: The lipid world. *Orig. Life Evol. Biospher.* (2001) 31: 119–145.

A.E. Smedman, I.B. Gustafsson, L.G. Berglund, B.O. Vessby, BO: Pentadecanoic acid in serum as a marker for intake of milk fat: relations between intake of milk fat and metabolic risk factors. *Amer. J. Clin. Nutr.* **69** (1999) 22–29.

J.W. Szostak, D.P. Bartel, P.L. Luisi: Synthesizing life. *Nature* 409 (2001) 387–390.

D. Voet, J. Voet: Biochemistry. 3rd edn. Wiley, New York 2004.

L. Yetukuri, M. Katajamaa, G. Medina-Gomez, T. Seppänen-Laakso, A. Vidal-Puig M. Oresic: Bioinformatics strategies for lipidomics analysis: characterization of obesity related hepatic steatosis. *BMC Syst. Biol.* 1 (2007) 12.